\documentclass{jetpl}
\usepackage[cp1251]{inputenc}
\usepackage{mathrsfs}
\usepackage{cite}

\twocolumn

\title{Stability of the coexistence phase of chiral superconductivity and noncollinear spin ordering with a nontrivial topology and strong electron correlations}

\rtitle{Stability of the coexistence phase of chiral superconductivity and noncollinear spin ordering...}

\sodtitle{Stability of the coexistence phase of chiral superconductivity and noncollinear spin ordering with a nontrivial topology and strong electron correlations}

\author{V.\,V. Val'kov$^{+}$, A.\,O. Zlotnikov\thanks{zlotn@iph.krasn.ru}$^{+}$}

\rauthor{V.\,V. Val'kov, A.\,O. Zlotnikov}

\sodauthor{Val'kov V.\,V., Zlotnikov A.\,O.}

\address{$^+$Kirensky Institute of Physics, Federal Research Center KSC SB RAS, 660036 Krasnoyarsk,
Russia}

\dates{05.03.2019}{19.04.2019}

\abstract{We show that the quantum charge and spin fluctuations,  while sufficiently renormalizing the magnetic order parameter, do not destroy the coexistence phase of chiral
$d+id$ superconductivity and $120^{\circ}$ spin ordering in a strongly correlated 2D system with a
triangular lattice. The nontrivial topology characterized by the topological invariant $\tilde{N}_3$ is also preserved.
It is shown that the Majorana mode exist among edge states in the topologically nontrivial phase. The spatial structure of such mode is determined. The spin and charge fluctuations shift the critical values of electron density at which quantum topological transitions occur. Increasing intersite Coulomb repulsion leads to decrease in the number of the topological transitions.}

\begin{document}

\maketitle

{\bf 1. Introduction}

Recently, several superconducting systems have been proposed in which the formation of Majorana edge states is possible. Among them are, for example, superconductors with chiral p-wave symmetry~\cite{Read_Green-00, Kitaev-01}, interfaces of a superconductor and a topological insulator~\cite {Fu_Kane-08, Snelder-15}, systems with spin-orbit interaction and proximity-induced superconductivity~\cite {Sau-10, Lutchyn-10, Sato-10, Val'kov-17a, Melnikov-18}. Experimentally, the greatest progress has been achieved for semiconductor nanowires InAs, InSb epitaxially coated by superconducting Al~\cite{Zhang-18}. A quantized peak of the zero-bias conductance has been detected upon increasing an external magnetic field~\cite{Zhang-18}.

Recently it has been found that the Majorana modes can be implemented
in materials with coexisting spin-singlet superconductivity and a long-range magnetic order~\cite{Martin-12,
Wang-13}. Such scenario is perspective due to the possibility of
appearance of the Majorana modes in a condensed matter system, while there is no the
spin-orbit interaction and an external
magnetic field. In this framework the existence of the
Majorana modes is demonstrated in superconducting systems with
helical magnetic ordering (such as, for example,
HoMo$_6$S$_8$, ErRh$_4$B$_4$)~\cite{Martin-12}.

It is widely believed that, due to bulk-boundary correspondence, the edge and Majorana modes appear when the ground state of the system with periodic boundary conditions corresponds to a phase with nontrivial topology.
The classification of such phases is carried out by using the topological invariant. In the simplest cases of 1D or 2D superconducting systems with broken time-reversal symmetry (symmetry class D~\cite{Schnyder-08}) described by quadratic Hamiltonians topological invariants are the $\mathbb{Z}_2$ invariant (Majorana number~\cite{Kitaev-01}) or $\mathbb{Z}$ invariant, respectively.
A relation between two invariants is described
in Ref.~\cite{Ghosh-10} for noncentrosymmetric superconductors. A transition between phases with different values of the topological invariant is implemented when the gap in the elementary excitation spectrum is closed~\cite{Volovik-03}.

Previously, it was predicted that the Majorana modes can be observed
in 2D systems with a triangular lattice in the coexistence phase
of chiral $d_1 + id_2$ superconductivity and a stripe magnetic order~\cite{Wang-13}. However, the further analysis
showed that chiral superconductivity does not
coexist with stripe spin ordering, but
coexists with a magnetic order corresponding to a $120^{\circ}$
structure~\cite{Val'kov-16}. The conditions for implementation of the Majorana modes in this coexistence phase are defined~\cite{Val'kov-18}. The
$\mathbb{Z}_2$ and $\mathbb{Z}$ invariants are calculated for the quadratic Hamiltonian and it is shown that topologically nontrivial phases
with an odd value of the $\mathbb{Z}$ invariant correspond to the parametric regions with the Majorana modes.

In the last time, the interest to the problem of electron correlations in topological phases has increased due to the fact that the correlations can lead to a change in the topological classification~\cite{Fidkowski-10}.
Although the classification is preserved for systems of even dimensions
described by the $Z$-invariant~\cite{Morimoto-15} (for example, the symmetry class D such as the above mentioned systems with the triangular lattice).

The topological classification of systems with interaction
can be based on a universal method, where the topological invariant
is determined in terms of the Green functions~\cite{Volovik-03}.
For the 2D systems with the gapped excitation spectrum
the $Z$-invariant is represented through the matrix Green function~\cite{Ishikawa-87}.
In~\cite{Volovik-03}, this invariant is denoted as $\tilde{N_3}$.
Use of $\tilde{N_3}$ has allowed to demonstrate a nontrivial topology
in quantum-Hall-effect systems~\cite{Ishikawa-87}, the phases of liquid helium $^3$He-A~\cite {Volovik-89},
$^3$He-B~\cite{Volovik-09}. Recently, it has been shown~\cite{Wang_Zhang-12} that the Green function at zero frequency
can be used to describe topological phases.

The development of a topological classification method of strongly
correlated materials with a triangular lattice
in the coexistence phase of chiral superconductivity and
noncollinear spin ordering is associated with
the study of the stability of such phase a with regard to
charge and spin fluctuations. It is related with the fact that, due to the reduced dimension and
frustrated exchange interaction in the triangular lattice,
the role of quantum fluctuations greatly increases in the
mechanism of destruction of the ordered phase.

In this work, we show that the magnetic order parameter is strongly renormalized in the experimentally
studied range of doping of sodium cobaltates. However, the structure of the spin ordering is preserved.
For the coexistence phase of chiral superconductivity and 120$^{\circ}$ spin structure the Green functions
are obtained and the topological characteristics of this phase are determined using the invariant $\tilde{N_3}$.
The change in the topological properties with increasing
electron concentration and Coulomb repulsion is established.
Namely, we identify a decrease in the number of topological transitions
with increasing the Coulomb interaction parameter.
Based on the solution of the equations for the
Green functions with open boundary condition along one direction of the
2D lattice the spatial structure of the Majorana
mode is demonstrated in the topologically nontrivial phase.

{\bf 2. Model}

To study the coexistence phase of chiral superconductivity and 120$^{\circ}$ spin ordering with
regard to spin-charge fluctuations and strong electron correlations we use $t-J-V$ model.
For definiteness, we consider electron-doped systems such as the superconducting Na$_x$CoO$_2$ hydrate~\cite{Baskaran-03}.
The Hamiltonian in the atomic representation is determined by the expression:
\begin{eqnarray}
\label{Ham}
H &=& \sum_{f \sigma} \left( \varepsilon - \mu \right) X_f^{\sigma \sigma} + \sum_{f} \left( 2\varepsilon + U - 2\mu \right) X_f^{2 2} +
\nonumber \\
&+& \sum_{fm\sigma} t_{fm} X_f^{2 \bar{\sigma}} X_m^{\bar{\sigma} 2} + \frac{V}{2} \sum_{f \delta} n_f n_{f+\delta} +
\nonumber \\
&+&  \sum_{f m} J_{fm} \left( X_f^{\uparrow\downarrow} X_m^{\downarrow \uparrow} -
X_f^{\uparrow \uparrow} X_m^{\downarrow \downarrow} \right),
\end{eqnarray}
where $\varepsilon$ is the bare electron energy, $\mu$ is the chemical potential, $U$ is the on-site Coulomb repulsion parameter, $t_{fm}$ is the hopping parameter, $V$ denotes the inter-site Coulomb interaction parameter, $n_f = X_f^{\uparrow \uparrow} + X_f^{\downarrow \downarrow} + 2X_f^{22}$ is the electron number operator at the site, and $J_{fm}$ is the exchange interaction parameter.

{\bf 3. Gapless excitations in the coexistence phase}

It is known~\cite{Volovik-03} that the topological transitions occur when the
elementary fermion excitations for the system with periodic boundary conditions
become gapless. The excitation spectrum in the noncollinear magnetic phase is
gapless on the Fermi contour (the line in the 2D Brillouin zone) at all levels
of doping. In the superconducting phase with the chiral $d_1 + id_2$ symmetry of the order parameter
the gapless excitations are realized only at the specific points of the Brillouin zone
(nodal points) with position depending on the electron density~\cite{Val'kov-15}.

It is not difficult to establish the conditions for nodal excitations
in the coexistence phase by analyzing the fermion spectrum~\cite{Val'kov-17}:
\begin{eqnarray}
\label{spectr}
E_{1,2 \, p} & = & \left[ \frac{1}{2} \left( \xi_{p}^2 + \xi_{p-Q}^2 + |\Delta_p|^2 + |\Delta_{-p+Q}|^2  \right) \right. +
\nonumber \\
& + & \left. R_{p}R_{p-Q}  \mp \lambda_{p} \right]^{1/2},
\end{eqnarray}
where
\begin{eqnarray}
\lambda_{p} = \left\{ \frac{1}{4} \left( \xi_{p}^2 - \xi_{p-Q}^2 + |\Delta_p|^2 - |\Delta_{-p+Q}|^2 \right)^{2} + \right.
\nonumber \\
\left. + R_{p}R_{p-Q} \left[ \left( \xi_{p} + \xi_{p-Q} \right)^2 + |\Delta_p+\Delta_{-p+Q}|^2 \right] \right\}^{1/2}.
\nonumber
\end{eqnarray}
The following notations are introduced $\xi_{p} = \varepsilon + U
-\mu+J_0(1-n/2)+V_0n+nt_{p}/2 = \xi_0+nt_{p}/2$, $n = \left\langle
n_f \right\rangle$ is the on-site electron density, $J_0$, $J_{Q}$ are the Fourier transforms of the exchange interaction for the quasi-momentums
$(0, 0)$, ${\bf Q}$, $V_0
= 6V$, $R_{p} = M(t_{p} - J_{Q})$, $R_{p - Q} = M(t_{p - Q} -
J_{Q})$, $M$ is the amplitude of the nonuniform magnetic order parameter determining the spin structure as
$\left\langle {\bf S}_f \right\rangle = M \left( \cos({\bf Q f}),
-\sin({\bf Q f}), 0 \right)$, $\Delta_p$ is the superconducting order parameter with the chiral $d_1+id_2$ and $p_1+ip_2$ invariants.

The nodal points of the spectrum (\ref{spectr}) in the coexistence phase are determined by the equations:
\begin{eqnarray}
\label{eq_gapless1}
\text{Im} \left( \Delta_p \Delta_{-p+Q}^* \right) & = & 0, \\
\label{eq_gapless2}
\left| \xi_p \Delta_{-p+Q} - \xi_{p-Q} \Delta_p \right| & = & 0, \\
\label{eq_gapless3}
R_{p}R_{p - Q} - \xi_p \xi_{p-Q} - \text{Re}\left( \Delta_p \Delta_{-p+Q}^* \right) & = & 0.
\end{eqnarray}

The spin 120$^{\circ}$ ordering on the triangular lattice
is defined by the vector ${\bf Q} = (2\pi/3, 2\pi/3)$. Hereinafter
coordinates of wave vectors are given in the basis of reciprocal lattice unit vectors. In this case, the equality
$\Delta_p = \Delta_{-p + Q} = 0$ satisfying Eqs. \eqref{eq_gapless1} and \eqref{eq_gapless2} holds at the center of the hexagonal Brillouin zone (point $\Gamma = (0,0)$) and at its boundaries (points $K = Q$, $K' = -Q$).
The parameters for the gapless excitations at the point $K'$ are found from the equation:
\begin{equation}
\label{Mn1}
M(n) = \left| \frac{-\tilde{\mu}(n,M) - 3 n/2(t_1 - 2t_2 +t_3)}{- 3(t_1 - 2t_2 +t_3) + 3(J_1 - 2 J_2)} \right|,
\end{equation}
where $\tilde{\mu}(n,M) = \mu(n,M) - J_0(1-n/2)-V_0n$;
$t_1$, $t_2$, $t_3$ are hopping parameters for three coordination spheres, $J_1$ and $J_2$ are the exchange parameters between nearest and next-nearest spins, respectively.

The excitation spectrum is gapless at the points $\Gamma$ and $K$ simultaneously
if the following equation is satisfied:
\begin{eqnarray}
\label{Mn2}
M^2(n) & - & \left( \frac{-\tilde{\mu}(n,M) - 3 n/2(t_1 - 2t_2 +t_3) }{ - 3(t_1 - 2t_2 +t_3) + 3(J_1 - 2 J_2) } \right) \times
\nonumber \\
&& \times \left( \frac{-\tilde{\mu}(n,M) + 3 n(t_1 + t_2 + t_3) }{ 6(t_1 + t_2 + t_3) + 3(J_1 - 2 J_2) } \right) = 0.
\end{eqnarray}

To find the exact conditions for development of the gapless spectrum of
fermion quasiparticles, and hence the conditions for
topological transitions with changing electron density
it is necessary to obtain the
expression for $M$ renormalized by
spin and charge fluctuations. This problem is solved in the
next section.

{\bf 4. Renormalization of the magnetic amplitude $\bf{M}$}

The theoretical description of magnets with the 120$^{\circ}$
ordering of localized spins is carried out in~\cite{Chubukov-94, Dzebisashvili-18,
Capriotti-99, White-07, Barabanov-92}.
The electronic ensemble on the triangular lattice is studied in the framework of the Hubbard model with mean-field~\cite{Pasrija-16} and slave-boson~\cite{Jiang-14} approximations. The phase diagrams with different spin and charge orderings have been obtained.
It is significant that the ground state with
120$^{\circ}$ spin ordering is preserved upon doping
near half filling. Using the Monte Carlo method, coexistence of
superconductivity and 120$^{\circ}$ spin ordering has been found near $n = 1.1$~\cite{Weber-06}.

For simplicity of derivation of the renormalizing $M$ the unitary transformation
of the Hamiltonian is made
\begin{eqnarray}
H \to \tilde{H} & = & U H U^{\dag}, \\ \nonumber U & = & \prod_f
\left[ \exp\left(i\frac{\pi}{2}S_f^y\right) \exp\left(i \theta_f
S_f^z \right) \right], \, \, \, \theta_f = -{\bf Q f},
\end{eqnarray}
corresponding to the rotation of the coordinate frame such as the
$z$ axis becomes aligned with $\left\langle
{\bf S}_f \right\rangle$ at each site.
The transformation rules for the operators are:
\begin{eqnarray}
S_f^x \to \tilde{S}_f^x & = & \cos(\theta_f )S_f^z - \sin(\theta_f
)S_f^y,
\nonumber \\
S_f^y \to \tilde{S}_f^y & = & \cos(\theta_f )S_f^y + \sin(\theta_f
)S_f^z,
\nonumber \\
S_f^z \to \tilde{S}_f^z & = & -S_f^x,
\nonumber \\
X_f^{\sigma \sigma} \to \tilde{X}_f^{\sigma \sigma} & = & \sum_{s
= \uparrow, \downarrow} X_f^{ss}/2 - \eta_{\sigma} \left(
X_f^{\uparrow \downarrow} + X_f^{\downarrow \uparrow} \right)/2,
\nonumber
\end{eqnarray}
\begin{eqnarray}
X_f^{\bar{\sigma} 2} \to \tilde{X}_f^{\bar{\sigma} 2} =
\exp(i\eta_{\bar{\sigma}}\theta_f/2)\left( X_f^{\bar{\sigma} 2} -
\eta_{\bar{\sigma}}X_f^{\sigma 2} \right)/\sqrt{2},
\end{eqnarray}
where $X_f^{\uparrow \downarrow} = S_f^x + iS_f^y$, $\eta_{\sigma} = +1, \, -1$, for $\sigma = \uparrow, \,
\downarrow$, respectively.

The transformed Hamiltonian with the obtained mean-field contributions has a form
\begin{eqnarray}
\label{tildeH} \tilde{H} &=& \sum_{f \sigma} \tilde{\xi}_{\sigma}
X_f^{\sigma \sigma} + \sum_{f} \left( 2\varepsilon + U + 2nV_0 -
2\mu \right) X_f^{2 2} +
\nonumber \\
&+& \sum_{fm\sigma} t_{fm} \cos \frac{{\bf Q}}{2} \left( {\bf f} -
{\bf m} \right) X_f^{2 \bar{\sigma}} X_m^{\bar{\sigma} 2} -
\nonumber \\
&-& i\sum_{fm\sigma} t_{fm} \sin \frac{{\bf Q}}{2} \left( {\bf f}
-  {\bf m} \right) X_f^{2 \sigma} X_m^{\bar{\sigma} 2} +
\\
&+& \sum_{f m} \frac{J_{fm}}{2} \left(1 +  \cos {\bf Q} \left(
{\bf f} -  {\bf m} \right) \right) X_f^{\uparrow\downarrow}
X_m^{\downarrow \uparrow}
\nonumber \\
&+& \sum_{f m} \frac{J_{fm}}{4} \left(1 - \cos {\bf Q} ( {\bf f} -
{\bf m} ) \right) \left( X_f^{\uparrow \downarrow} X_m^{\uparrow
\downarrow} + X_f^{\downarrow \uparrow} X_m^{\downarrow \uparrow}
\right).
\nonumber
\end{eqnarray}
Only those terms describing the exchange interaction are taken into account
that lead to fluctuation-induced corrections in the one-loop
approximation. Here $\tilde{\xi}_{\sigma} = \varepsilon - \mu -
(1-n/2)J_0 + nV_0 - \eta_{\sigma}h_Q$ and $h_Q = -MJ_Q$.

\begin{figure}[htb!]
\begin{center}
\includegraphics[width=0.35\textwidth]{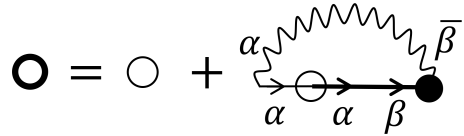}
\caption{Fig.~1 Diagrams for $\left\langle X_f^{\uparrow\uparrow} \right\rangle$ }
\end{center}
\end{figure}

Using the completeness condition of single-ion states and the definition connecting the electron concentration
to filling numbers one can obtain the
expression for the amplitude of the magnetic parameter order which is convenient
to calculate: $M = n/2 - 1 + N_{\uparrow}$. Here $N_{\uparrow} = \left\langle X_f^{\uparrow\uparrow} \right\rangle$ is the filling number of a state with a spin projection $1/2$. To find this number, the diagrammatic form of  perturbation theory in the atomic representation is used. The diagrammatic representation for $N_{\uparrow}$ with regard to the first contributions caused by the spin and charge fluctuations is shown in Fig. 1. In the second diagram,
$\alpha$ denotes the type of elementary excitation
(root vector \cite{Zaitsev1975, Zaitsev1976}).
For $\alpha=(\uparrow~\downarrow)$
this diagram determines contributions from spin fluctuations. In this case
the second root vector takes two values $\beta=(\uparrow~\downarrow),~(\downarrow~\uparrow)$
 \cite{Zaitsev1976, Val'kov-01}. The charge
fluctuations are described by two terms in accordance with the fact that
the fermion root vector is $\alpha = (\uparrow~2)$ and
$\beta$ takes two values $(\uparrow~2)$ and $(\downarrow~2)$.
According to the diagrammatic technique rules~\cite{Zaitsev1975, Zaitsev1976,
Val'kov-01} we get the expression for $M$ in the limit $T \to 0$:
\begin{eqnarray}
\label{eq_M} && M(n) = \frac{n}{2} + \frac{1}{2} - \sum_q \frac{A_q^+/2 -
J_Q}{2\gamma_q} -
\frac{1}{2} \sum_p \left( f_{1p} + f_{2p} \right) -
\nonumber \\
& - &  M \sum_{p} \frac{J_Q-t_p^+}{\varepsilon_{2p} - \varepsilon_{1p}} \left( f_{1p} - f_{2p} \right),
\end{eqnarray}
where $A_{q}^{+} = J_{q} + \left( J_{q-Q} + J_{q + Q}
\right)/2$, $\gamma_{q}$ is connected with the spectrum of spin-wave excitations as
\begin{equation*}
\omega_{0 q} = 2M \gamma_{q} = 2M\sqrt{\left( J_{q} - J_Q \right)
\left[ \frac{J_{q - Q} + J_{q + Q}}{2} - J_Q \right]},
\end{equation*}
$f_{jp} \equiv f(\varepsilon_{jp}/T)$ are the Fermi-Dirac functions.
The branches of the fermion spectrum are expressed as:
\begin{eqnarray} \varepsilon_{1,2p} = \xi_0 + nt_{p}^{+}/2 \mp \sqrt{\left(nt_{p}^{-}/2 \right)^2 + R_{p - Q/2} R_{p + Q/2}} \nonumber
\end{eqnarray}
with $t_{p}^{\pm} = \left( t_{p - Q/2} \pm t_{p + Q/2} \right)/2$.

In the derivation of Eq.~\eqref{eq_M} it is essential that the average $\left\langle X_f^{22} \right\rangle$ is expressed through the electron Green functions $G_{\sigma 2,\sigma 2}(p, i\omega_n)$. This leads to the following equation for the chemical potential:
\begin{eqnarray}
\label{eq_n}n-1 & = & \frac{n}{4} \sum_p \left( f_{1p} + f_{2p}
\right) +
\nonumber \\
& + & M^2 \sum_{p} \frac{J_Q-t_p^+}{\varepsilon_{2p} -
\varepsilon_{1p}} \left( f_{1p} - f_{2p} \right).
\end{eqnarray}

A decrease in the magnetization connected with the
spin fluctuations does not depend on electron density as it follows from (\ref{eq_M}). At
half filling ($n=1$), when hoppings are prohibited,
the magnetization of the 120$^{\circ}$ structure is determined by the well-known
expression~\cite{Chubukov-94}. The antiferromagnetic exchange interaction between next nearest neighbors
with the parameter $J_2$ leads to frustrations and reduction in $M$.

\begin{figure}
\begin{center}
\includegraphics[width=0.4\textwidth]{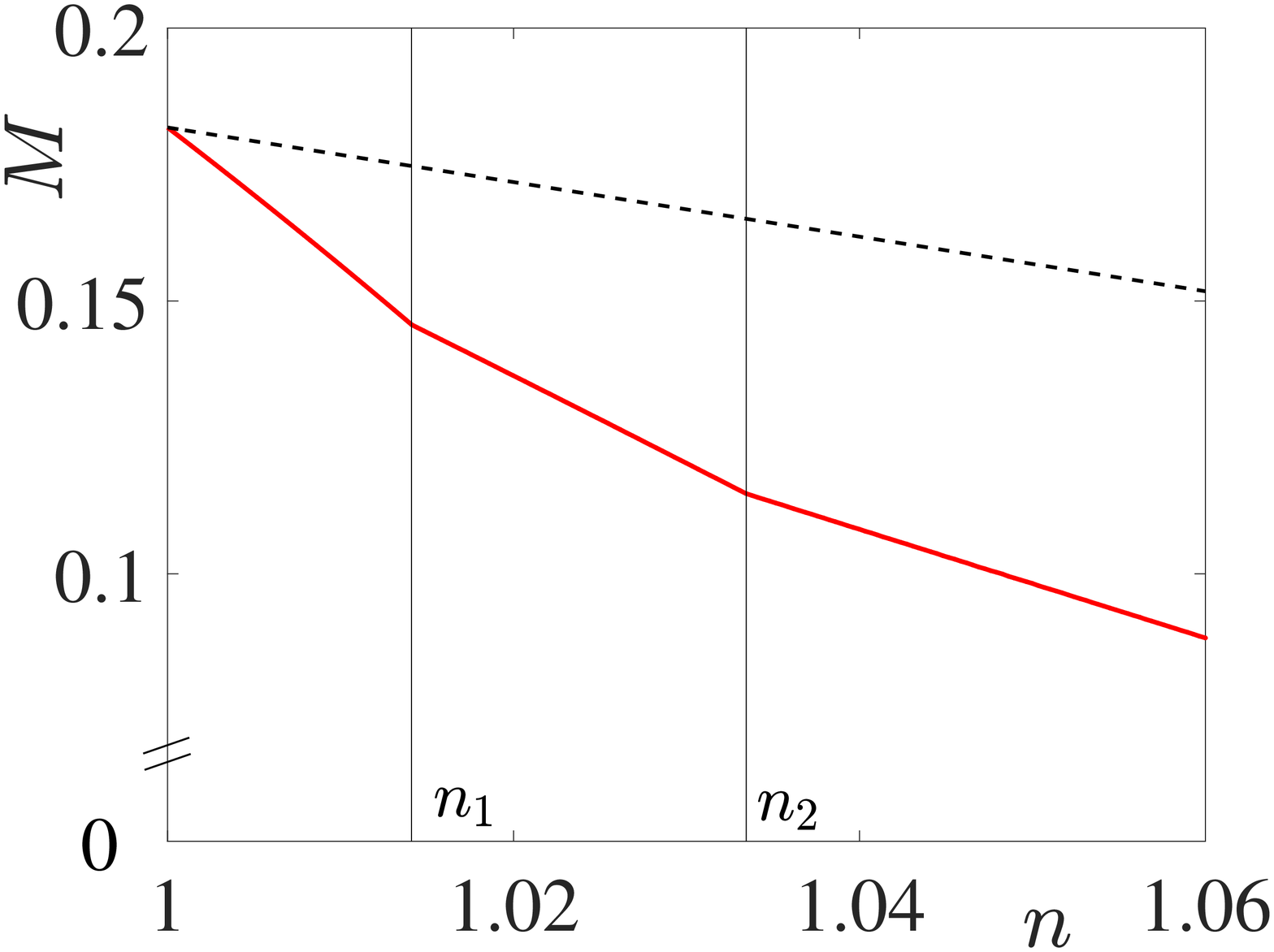}
\caption{Fig.~2 The dependence of the magnetization $M$ on the electron density (solid line) at $J_1 = 0.5 t_1$, $J_2 = 0.02 t_1$ as compared
to (dashed line) the result obtained disregarding
charge fluctuations. The vertical lines denote the critical concentrations $n_{1}$ and
$n_{2}$ for topological transitions in the coexistence phase}
\end{center}
\end{figure}

The density dependence of the magnetization at the parameters $J_1 =
0.5 t_1$, $J_2 = 0.02 t_1$ is shown in Fig. 2 by the solid line.
It is seen that hoppings near half filling
lead to a stronger decrease in the magnetization with increasing
density in comparison with
the trivial result $1-n/2$ (dashed line in Fig. 2 is obtained taking into account this result and the contribution from spin fluctuations) in the vicinity of half filling.
The vertical lines denote the densities $n_1
= 1.014$ and $ n_2 = 1.033 $ at which
gapless excitations are realized in the coexistence phase. The features of
these densities are also manifested in the energy
spectrum of fermion states for noncollinear spin
ordering. At densities $1<n<n_1$ the states are filled near the point $ K^{\prime}$ of the Brillouin zone, as it can be seen from Fig. 3.
At densities $n_1<n<n_2$
the filling of states near the $ \Gamma $ and $ K $ points occurs (see
Fig.~4). Due to such density evolution
the arising corrections to the magnetization lead to a kink in the dependence $ M(n) $ at
$n = n_1$. Above the density $ n_2 $ the upper band with the minimum at the $ K^{\prime} $ point  starts to be filled. As a result such processes lead to a a decrease in the slope of the dependence $ M(n) $ and
the occurrence of the second kink at $ n = n_2 $. The spectrum of the upper band is described by the
expression $ \varepsilon_{2p} $.

\begin{figure}
\begin{center}
\includegraphics[width=0.35\textwidth]{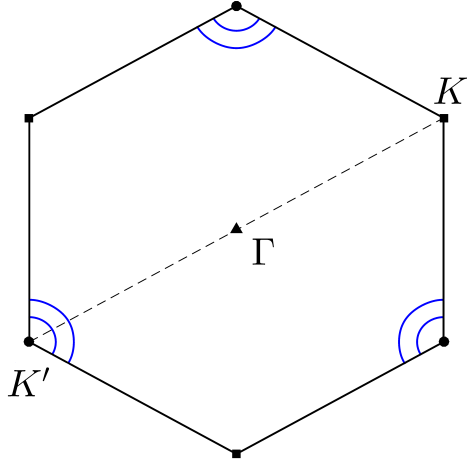}
\caption{Fig.~3 Fermi contours for densities $1 < n < n_1$,
$n_{11} = 1.005$, $n_{12} = 1.013$}
\end{center}
\end{figure}

\begin{figure}
\begin{center}
\includegraphics[width=0.35\textwidth]{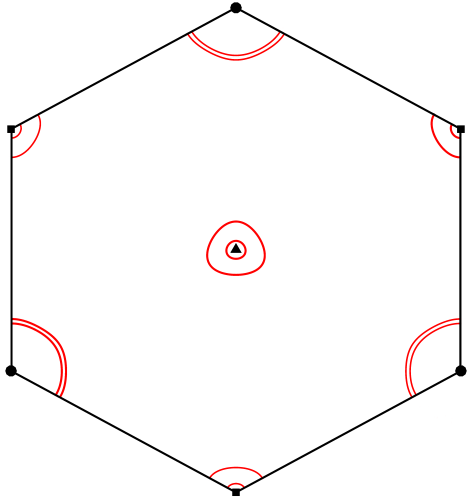}
\caption{Fig~4 Fermi contours for densities $n_1 < n < n_2$, $n_{21} = 1.015$, $n_{22} = 1.025$}
\end{center}
\end{figure}

The fermion spectrum is shown in Figs. 5 and 6 for different values of the density $n = n_1$ and $n = n_2$, respectively.
The spectrum indicates realization of gapless excitations in the coexistence phase. The reported effects argue that
the existence of gapless excitations and, accordingly,
topological transitions can be revealed by the behavior of the magnetization and related characteristics.

The density dependencies of the amplitude $\Delta_{21}$ describing the superconducting pairings due to the exchange interaction with the first coordination sphere are shown in Fig. 7.
The relevant self-consistent equations are given in~\cite{Val'kov-17}.
In the next section, we show that a topological transition with
a change in the topological invariant occurs when the gapless
excitations are implemented in the coexistence phase of superconductivity and
120$^{\circ}$ ordering.

\begin{figure}
\begin{center}
\includegraphics[width=0.45\textwidth]{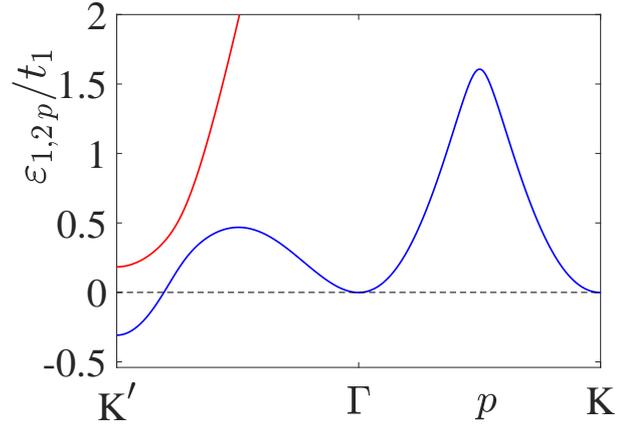}
\caption{Fig.~5 Fermi spectrum with regard to the noncollinear spin ordering along the $K^{\prime}$-$\Gamma$-$K$ direction of the Brillouin zone for the density $n = n_1$. The energy is counted from the chemical potential}
\end{center}
\end{figure}

\begin{figure}
\begin{center}
\includegraphics[width=0.45\textwidth]{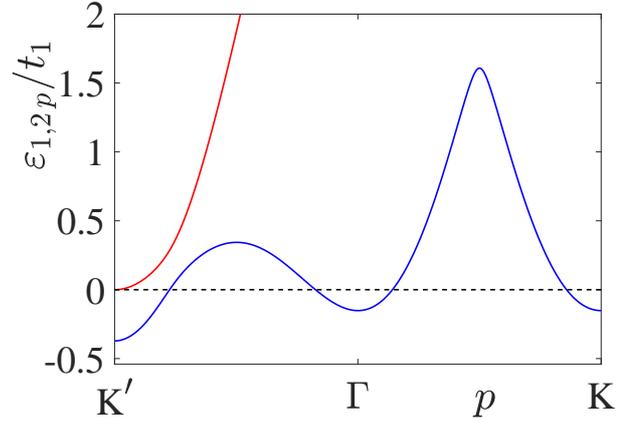}
\caption{Fig.~6 The Fermi spectrum for the density $n = n_2$}
\end{center}
\end{figure}

{\bf 5. Topological invariant $\tilde{N}_3$ and Majorana modes}

To solve the problem of a nontrivial topology of the
coexistence phase of superconductivity and noncollinear
magnetism at strong electron correlations, we
use the method based on the analysis of the integer-valued
topological invariant
$\tilde{N}_3$~\cite{Volovik-03}:
\begin{eqnarray}
\label{Z_inv} \tilde{N}_{3}& = &
\frac{\varepsilon_{\mu\nu\lambda}}{24\pi^{2}} \times
\\
&\times&
\int\limits_{-\infty}^{\infty}d\omega\int\limits_{-\pi}^{\pi}
dk_{1}dk_{2}
\text{Tr}\left(\widehat{G}\partial_{\mu}\widehat{G}^{-1}
\widehat{G}\partial_{\nu}\widehat{G}^{-1}
\widehat{G}\partial_{\lambda}\widehat{G}^{-1} \right). \nonumber
\end{eqnarray}
Here, the repeated indices $\mu$, $\nu$, $\lambda = 1, \, 2, \, 3$
imply summation,  $\varepsilon_{\mu\nu\lambda}$ is the Levi-Civita symbol, $\partial_{1} \equiv \partial/\partial k_1$,
$\partial_{2} \equiv \partial/\partial k_2$, $\partial_{3} \equiv
\partial/\partial \omega$, and $\widehat{G}(i\omega,k)$ is the matrix
Green’s function whose poles determine the spectrum
of elementary fermion excitations
(details are presented in the supplementary material).

The value $\tilde{N}_3 = 0$ corresponds to the topologically
trivial phase. In a topologically nontrivial phase, $\tilde{N}_3 \ne 0$.
Transitions between phases with different $\tilde{N}_3$
values are topological transitions.

\begin{figure}[htb!]
\begin{center}
\includegraphics[width=0.4\textwidth]{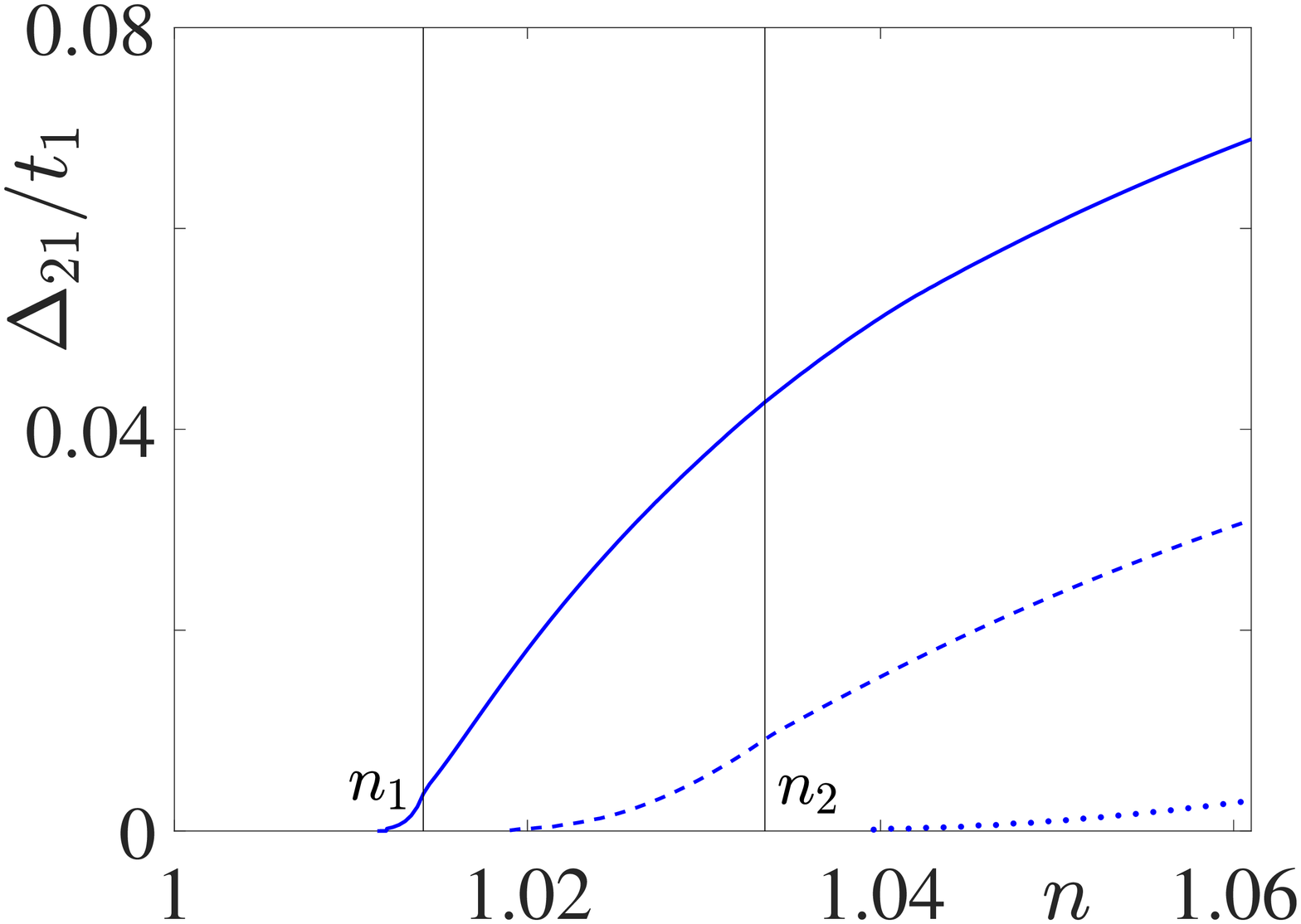}
\caption{Fig.~7 The dependence of the amplitude $\Delta_{21}$ on density for different values of the parameter of intersite Coulomb repulsion: $V = 0$
(solid line), $V = 0.3 t_1$ (dashed line), $V = 0.6 t_1$
(dotted line)}
\end{center}
\end{figure}

The calculation of the number $\tilde{N_3}$ shows that the
coexistence phase of superconductivity and 120$^{\circ}$ spin
ordering is topologically nontrivial with $\tilde{N_3}
\ne 0$. The coexistence phase at $V = 0$ occurs in a fairly wide density
range (see Fig. 7), but a particular $\tilde{N_3}$ value can be
different. The sequence of changes in $\tilde{N_3}$ upon an increase
in the density of fermions is as follows:
\begin{eqnarray}
(\tilde{N_3}=-1)~~{\stackrel{n = n_{1}}{\longrightarrow}}~~
(\tilde{N_3} = 3)~~{\stackrel{n = n_{2}}{\longrightarrow}}~~
(\tilde{N_3} = 2).
\end{eqnarray}
Topological transitions occur at the same parameters
at which the bulk spectrum of elementary excitations
becomes gapless. It is substantial that these conditions
for existence of topological transitions are independent
of the magnitude of the superconducting order
parameter.

As seen in Fig. 7, different numbers of topological
transitions occur in the coexistence phase depending
on the parameter $V$. Indeed, at $V = 0$, two such transitions
occur at the densities $n_{1}$ and $n_{2}$; as $V$ increases to
$V = 0.3 t_1$, the transition at $n=n_{1}$ disappears, but the
topological transition at the density $n_{2}$ holds; and
topological transitions are absent at $V = 0.6
t_1$.

To determine the structure of the Majorana mode,
we use a method similar to that used for models disregarding interactions~\cite{Val'kov-18}. We consider a system with
the triangular lattice containing a finite number ($N_1$)
of sites along the direction of the translation vector ${\bf a}_1$,
whereas periodic boundary conditions are imposed
along the ${\bf a}_2$ direction (cylindrical geometry). According
to the solution of the system of equations in the
coordinate–momentum representation, the low-energy
``quasiparticle'' Green’s function can be represented
in the form
\begin{eqnarray}
\label{GRFM} \left( i \omega_m - \varepsilon_{j k_2} \right)
G_{\alpha_j, \downarrow 2} \left( k_2; n'; i\omega_m \right) =
\left( S^{\dag} \right)_{jn'},
\end{eqnarray}
where $\varepsilon_{j k_2}$ are the branches of the excitation spectrum
with $j = 1,2, \dots N_1$, and $S$ is the transformation matrix diagonalizing
the initial matrix of the system of equations.

The relation between the Green’s function found
from Eq. (\ref{GRFM}) and initial Green’s functions in the
coordinate–momentum representation makes it possible
to determine the operators of elementary excitations
for the coexistence phase in the cylindrical
geometry in terms of the Hubbard fermion operators:
\begin{eqnarray}
\label{DEFUV} \alpha_{j k_2} & = & \sum_{l=1}^{N_1} u_{jl}
X_{k_2,l,\uparrow} + w_{jl} X_{k_2-Q_2,l,\downarrow}
\nonumber \\
& + & z_{jl} X^{\dag}_{-k_2+Q_2,l,\uparrow} + v_{jl}
X^{\dag}_{-k_2,l,\downarrow}.
\end{eqnarray}
According to this definition and symmetry reasons~\cite{Val'kov-18}, the Majorana mode in the cylindrical geometry
occurs at $K_2 = -K_2+Q_2+G$, i.e., at $K_2 = -Q_2 = -2\pi/3$, when the excitation energy is zero, as seen in Fig. 8.

\begin{figure}
\begin{center}
\includegraphics[width=0.4\textwidth]{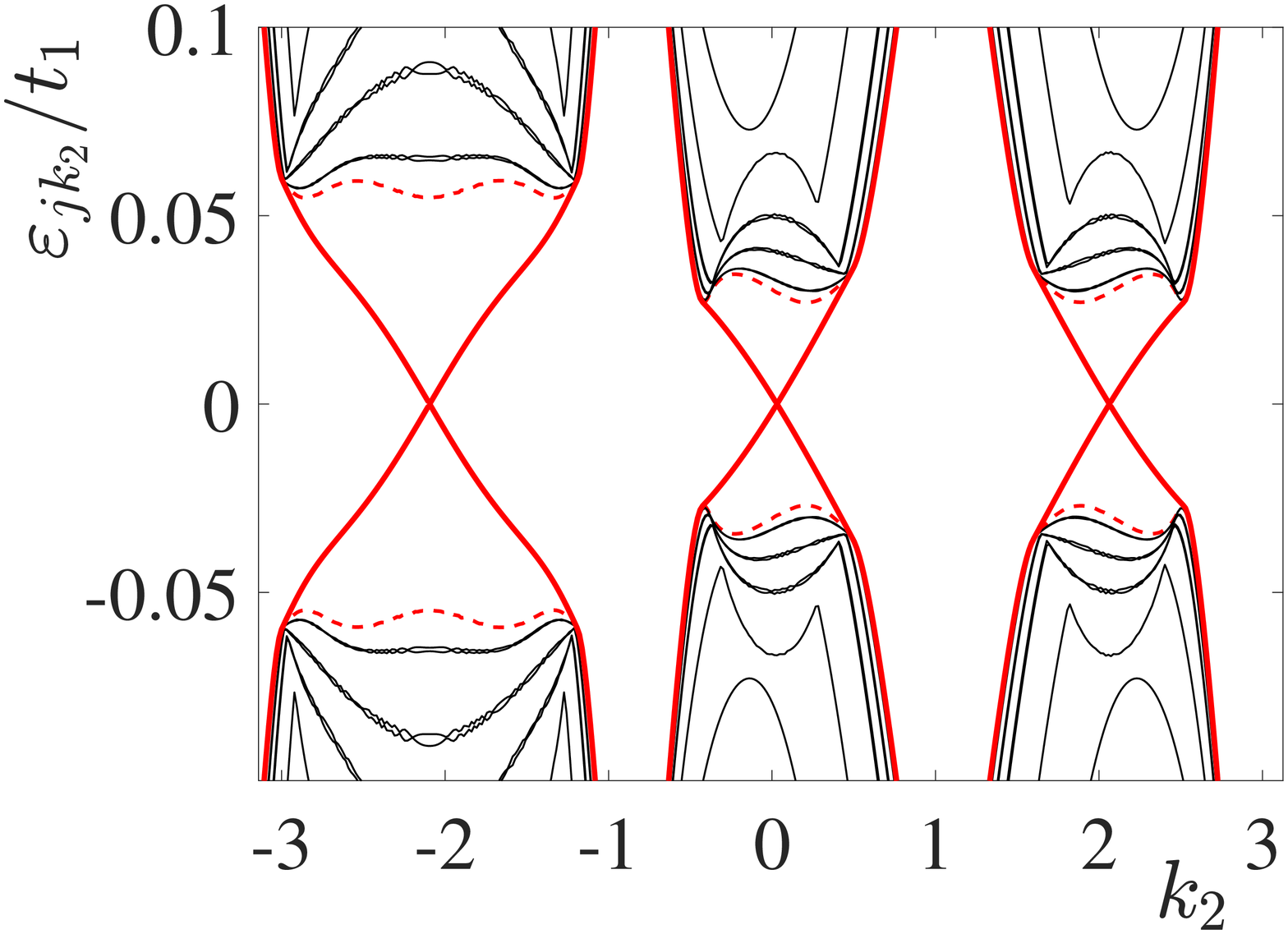}
\caption{Fig.~8 Spectrum of fermion excitations in
the coexistence phase for the cylindrical geometry}
\end{center}
\end{figure}

The dependences of several branches of the excitation
spectrum on the quasimomentum
$k_2$ at the density $n = 1.025$ and $V = 0$ are shown in
Fig. 8, where the other parameters are the same as
those used for Figs. 2 and 7. The dashed line denotes the
boundary of the bulk excitation spectrum at periodic
boundary conditions along both directions of the triangular lattice. It is seen that the excitation spectrum
in this case has an energy gap. Edge states appear inside the spectrum gap (shown by thick solid line) in the cylindrical geometry. Thin solid lines selectively show the branches lying in the region of the bulk spectrum.

For the visualization of the spatial structure of the
Majorana mode ($K_2 = - Q_2$), we use the Kitaev
approach. To this end, we introduce two Hermitian
operators $b^{\prime} =
\alpha_1 + \alpha_1^{\dag}$ and $b^{\prime \prime} =
i(\alpha_1^{\dag}-\alpha_1)$. Then,
using expansion (\ref{DEFUV}), we express these operators in
terms of Majorana operators in the atomic representation
\begin{eqnarray}
\gamma_{A l \sigma} = X_{l \sigma} + X^{\dag}_{l \sigma},~~~
\gamma_{B l \sigma} = i\left(X^{\dag}_{l \sigma}-  X_{l
\sigma}\right).
\end{eqnarray}
These transformations give
\begin{eqnarray}
\label{DEF_b_prim} b^{\prime}=\sum_{l=1}^{N_1}
\left\{Re(u_{1l}+z_{1l})\gamma_{A l \uparrow}
+Re(w_{1l}+v_{1l})\gamma_{A l \downarrow}\right\}-
\nonumber \\
\sum_{l=1}^{N_1} \left\{Im(u_{1l}-z_{1l})\gamma_{B l \uparrow}
+Im(w_{1l}-v_{1l})\gamma_{B l \downarrow}\right\},\nonumber
\end{eqnarray}
\begin{eqnarray}
\label{DEF_b_pr_pr} b^{\prime\prime}=\sum_{l=1}^{N_1}
\left\{Im(u_{1l}+z_{1l})\gamma_{A l \uparrow}
+Im(w_{1l}+v_{1l})\gamma_{A l \downarrow}\right\}+
\nonumber \\
\sum_{l=1}^{N_1} \left\{Re(u_{1l}-z_{1l})\gamma_{B l \uparrow}
+Re(w_{1l}-v_{1l})\gamma_{B l \downarrow}\right\}\nonumber
\end{eqnarray}
Figure 9 shows the dependence of the coefficients $A_l=Re(u_{1l}+z_{1l})$ and $B_l=Im(u_{1l}+z_{1l})$ of the decompositions
of the operators $b^{\prime}$ and $b^{\prime \prime}$ in Majorana operators
in the atomic representation on the site number. It is
seen that this dependence is localized near different
edges. Other expansion coefficients demonstrate a
similar localization and, for this reason, are not
shown.

\begin{figure}
\begin{center}
\includegraphics[width=0.48\textwidth]{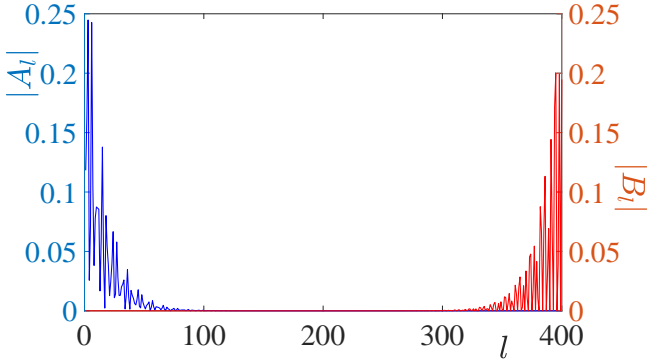}
\caption{Fig.~9 Coefficients $A_l = \text{Re}(u_{1l} + z_{1l})$ and
$B_l = \text{Im}(u_{1l} + z_{1l})$ versus the site number for the density
$n = 1.025$ and $N_1 = 400$}
\end{center}
\end{figure}

After the topological transition to the region with
$\tilde{N}_3 = 2$ for densities $n > n_{2}$, a gap for the branch of edge
states opens in the excitation spectrum of the finite
system at $K_2 = -Q_2$ and the Majorana mode is absent.
The Majorana mode also occurs at $K_2 = -Q_2$ in the
coexistence phase with $N_3 = -1$ in the narrow density
range $n < n_{1}$. In this region, in contrast to the region
with $N_3 = 3$, the branch of edge states is formed only
near $K_2 = -Q_2$. However, such topological phase is of
low practical interest because the superconducting gap
is very narrow even disregarding the intersite Coulomb
interaction.

{\bf 6. Conclusions}

The results obtained in this work show that strong
electron correlations significantly renormalize the
spin structure parameter but do not destroy the coexistence
phase of chiral superconductivity and noncollinear
magnetic ordering. It has been found that a
nontrivial topology also holds in this case, which is
important for the formation of Majorana modes in this
phase at open boundary conditions. The nontrivial
topology has been proven using the topological invariant $\tilde{N_3}$
calculated in terms of the Green’s functions
determined by means of the Hubbard operators. The
atomic representation has made it possible not only to
correctly describe the effects of strong electron correlations
but also to study the structure of the Majorana
mode for the strongly correlated coexistence
phase of chiral superconductivity and noncollinear
spin ordering by introducing Majorana operators in
the atomic representation. The diagrammatic technique
for the Hubbard operators has allowed the calculation
of contributions from spin and charge fluctuations
to the macroscopic characteristic of the magnetic
structure. Particular calculations near half filling
within the $t-J-V$ model have demonstrated a change
in the topological invariant $\tilde{N_3}$ at the variation of the
electron density. It has been found that the character
of a change in $\tilde{N_3}$ depends on the intersite Coulomb
interaction. An important conclusion has been made
from the form of the dependence of $\tilde{N_3}$: depending on
the model parameters and external conditions, either
edges states which are Majorana bound states or edge
states which do not belong to the Majorana type can
be formed in the coexistence phase. The transition
between these two regimes occurs as a quantum topological
transition in density.

We are grateful to M.S. Shustin for stimulating discussions.
This work was supported by the Russian
Foundation for Basic Research (project nos. 19-02-
00348-a and 18-32-00443-mol-a); jointly by the Russian
Foundation for Basic Research, the Government
of Krasnoyarsk Region, and the Krasnoyarsk Region Science and Technology Support Fund (project no. 18-42-243002
``Manifestation of Spin–Nematic Correlations in
Spectral Characteristics of the Electronic Structure
and Their Influence on Practical Properties of Cuprate Superconductors''); and by the Presidium of
the Russian Academy of Sciences (program no. I.12
``Fundamental Problems of High-Temperature Superconductivity'').
A.O.Z. acknowledges the support of
the Council of the President of the Russian Federation
for State Support of Young Scientists and Leading Scientific
Schools (project no. MK-3594.2018.2).

\emph{The section 5 and conclusions translated by R.~Tyapaev}

\end{document}